\documentclass[twocolumn,aps,prl]{revtex4-2}
\usepackage{graphicx}% Include figure files
\usepackage{dcolumn}% Align table columns on decimal point
\usepackage{bm}% bold math

\begin{document}

\title{Revealing the regularities of electron correlation energies associated with valence electrons
in atoms in the first three rows of the periodic table}

\author{G.-Q. Hai$^a$}
\email{Corresponding author: hai@ifsc.usp.br}
\author{L. C\^andido$^{b}$}
\author{B. G. A. Brito$^c$}
\author{Y. Liu$^d$ }

\affiliation{$^a$Instituto de F\'{\i}sica de S\~{a}o Carlos, Universidade de S\~{a}o Paulo, 13560-970, S\~{a}o Carlos, SP, Brazil}
\affiliation{$^b$Instituto de F\'{i}sica,  Universidade Federal de Goi\'{a}s, 74.001-970, Goi\^{a}nia, GO, Brazil}
\affiliation{$^c$Departamento de F\'isica, Universidade Federal do Tri\^{a}ngulo Mineiro, 38.025-180, Uberaba, MG, Brazil}
\affiliation{$^d$Research School of Chemistry, The Australian National University, ACT 2601, Australia}

%\date{\today}

\begin{abstract}
Electronic correlation is a complex many-body effect and the correlation energy depends on the specific
electronic structure and spatial distribution of electrons in each atom and molecule. Although the total
correlation energy in an atom can be decomposed into different components such as inter-orbital and
intra-orbital pair-correlation energies (PCE), it is generally believed that the PCEs in different atoms
cannot be the same. In this work, we investigate the correlation energies of the atoms in the first
three rows of the periodic table (He to Ar). It is found that when the correlation energy
is defined as the difference between the exact ground-state energy and the unrestricted Hartree-Fock (UHF) energy,
the inter- and intra-orbital PECs associated with the valence electrons of the atoms in the same row
of the periodic table have the same values. These PCEs are not entangled and their values depend only on
the electron orbitals. For two specific orbitals, the inter-orbital correlation energy is
the same between two electrons of parallel spins or anti-parallel spins.
We also show that the effects of orbital relaxation on the correlation energy are surprisingly small.
\end{abstract}

\maketitle

Electron correlation energy ($E_{\rm c}$) in atoms and molecules is defined as the difference between
the ground-state energy ($E_{\rm exact}$) and the Hartree-Fock energy ($E_{\rm HF}$) in the complete basis
set limit\cite{Lowdin,Pople,HFtheory,Levine}, i.e., $E_{\rm c}=E_{\rm exact} - E_{\rm HF}$,
where $E_{\rm exact}$ is obtained from the exact solution of the non-relativistic electronic
Schr\"{o}dinger equation with fixed nuclei. The Hartree-Fock (HF) energy contributes to more
than 99$\%$ of the total ground-state energy in most atoms and molecules. Although $E_{\rm c}$
is only a small fraction of the total energy, it is a quantity at the heart of chemistry
playing a crucial role in the determination of the physicochemical properties of atoms and molecules.
In order to obtain the correlation energy of a many-electron atom or a molecule,
one must perform sophisticated calculations to find out the total energy of each many-body system.
Although accurate knowledge of correlation energies is essential for understanding and predicting
the behavior of molecules as well as their thermochemical properties,
many aspects of electron correlation and the properties of the correlation energy
are still poorly understood.\cite{HFtheory,Levine,hattig,Brito14}

In the long history of investigation of the electron correlation, some regularities are
observed by examining trends in the correlation energy of atoms as we move across the periodic
table or within a specific group of elements.\cite{HFtheory,Levine,hattig,Brito14,OS61,Clementi,Davidson,Davidson93,Davidson96,McCarthy11}
The correlation energy is influenced by the specific electron configuration and the effective
nuclear charge experienced by the electrons. It varies significantly depending on the position
of the atom in the periodic table. Elements in the same group tend to have similar valence
electron configurations and exhibit similar correlation energy patterns. Generally, for neutral atoms,
$E_{\rm c}$ increases roughly linearly with increasing the atomic number but the
ratio $E_{\rm c}/E_{\rm {exact}}$ decreases. In dealing with the correlation energy in an atom,
it is possible to separate the correlation energy into different components such as inter-orbital
and intra-orbital pair-correlation energies (PCE). For different atoms, however, these PCEs
are unlikely to be the same for different atoms.
In this work, we attempt to establish some regularities in the electron correlation energies of the atoms.
We show for the first time that, there are simple relations for the PCEs associated with
the valence electrons of different atoms in the same period of the periodic table.
These PCEs are not entangled and their values depend only on the electron orbitals.
We have obtained very accurate values of the PCEs for atoms in the first three rows of the periodic table.
For example, all the PCEs associated with the valence electrons in the 2$p$ orbitals in B, C, N, O, F and Ne atoms
share the same values $\varepsilon_{2p\mbox{-}2p}=-7.3\pm 0.2$m$E_{\rm h}$ (for two electrons
in different 2$p$ orbitals), $\varepsilon_{2p\mbox{-}2s}=-3.9\pm 0.2$m$E_{\rm h}$
(for one electron in 2$p$ orbital and the other in 2$s$ orbital),
and $\varepsilon_{2p^2}=-35.5\pm 0.8$m$E_{\rm h}$ (for two electrons in the same 2$p$ orbital).
Our calculations show that the inter-orbital PCEs have the same value between two electrons with parallel spins
or anti-parallel spins. It means that the spin states do not {\em directly} affect the PCEs in atoms.
We will also show that the orbital relaxation effects on the correlation energy is very small.
These results provide valuable insights into the properties of electron correlation energy in atoms.

The effective nuclear potential experienced by the electrons in an atomic system
may strongly affect the correlation energy. For the same electron number and
configuration, the correlation energy in the ground state of an atomic ion is dependent
on the nuclear charge $Z$.\cite{Davidson,Gill2011} For example, the correlation energy $E_{\rm c}$
of the four-electron Be-like ion increases almost linearly with increasing $Z$. Therefore,
in order to facilitate our investigation to better understand the nature of
the electron correlation, we will focus on the correlation energies associated with the valence
electrons in the neutral state of the atoms. The correlation energies of the valence
electrons in the outermost shell will be analyzed with the neutral atomic state as a
reference. We will investigate the correlation-energy gain
$\Delta E_c(Z) = E^{(0)}_{\rm c}(Z) - E^{(+)}_{\rm c}(Z)$
in the formation of the neutral atom (with atomic number $Z$) by adding one valence
electron to its cation, where $E^{(0)}_{\rm c}(Z)$ and $E^{(+)}_{\rm c}(Z)$ are the correlation
energies of the neutral atom and the corresponding cation, respectively.
The inverse of the above process “ion + electron $\to $ neutral atom” is atomic
ionization and the ionization energy is experimentally measurable.
Analyzing the correlation energies associated to the valence electrons, we may minimize
the influences of the nuclear potential as well as the so-called exclusion effects.\cite{OS61,Ranasinghe}
In fact, the correlation energies of these valence electrons are of great
importance tending to be responsible for the element’s chemical properties.

The correlation energy $E_{\rm c}$ depends upon whether a restricted ($E_{\rm RHF}$) or unrestricted
Hartree-Fock energy ($E_{\rm UHF}$) is used in its definition.\cite{Lowdin,Pople,HFtheory,Levine,Gill,Gill2011}.
Thus, there are different correlation energies  $E_{\rm c(R)} = E_{\rm exact} - E_{\rm RHF}$
defined by Löwdin\cite{Lowdin} and $E_{\rm c(U)} = E_{\rm exact} - E_{\rm UHF}$ defined by Pople and Binkley\cite{Pople}.
The restricted Hartree-Fock (RHF) theory uses a single atomic or molecular orbital twice,
one multiplied by the $\alpha$ spin function and the other multiplied by the $\beta$ spin
function in the Slater determinant, then the resulting approximate wavefunction is
an eigenfunction of the spin-squared operator ${\bf S}^2$ and the spin component operator $S_z$,
simulating the corresponding properties of the exact wavefunction at the non-relativistic
limit.\cite{Lowdin,Pople,HFtheory,Levine}
A disadvantage of the RHF theory is manifested in its incorrect limit for molecule dissociation.
On the other hand, in the unrestricted Hartree–Fock (UHF) theory, the $\alpha$ and $\beta$
spin functions are assigned to two completely independent sets of spatial orbitals.
It has the advantage of dissociation to correct fragments for most molecular ground states.\cite{Pople}
The additional flexibility in UHF permits a lower limiting energy to be achieved but its total
wavefunction is not an eigenfunction of the total spin-squared operator ${\bf S}^2$.
It was pointed out by Pople and Binkley that, according to the variational theorem
$E_{\rm UHF}$ is lower than $E_{\rm RHF}$ because there are no symmetry constraints are placed
on the orbitals in the UHF theory.\cite{Pople} Therefore, $E_{\rm UHF}$ obtained within
the complete basis set limit is the limiting energy which can be reached within the mean-field theory.
Consequently, the relations $E_{\rm RHF} \ge E_{\rm UHF} > E_{\rm exact}$
and $|E_{\rm c(R)}| \ge |E_{\rm c(U)}|$ hold.

In the literature, most studies of the electron correlation energies in atoms
are about the correlation energy $E_{\rm c(R)}$ with the RHF energy as the reference.
Different correlation energies $E_{\rm c(U)}$ and $E_{\rm c(R)}$ were calculated and compared
for many atoms and molecules\cite{Gill} by O'Neill and Gill. It was pointed out from the point
of view of static and dynamical electron correlation\cite{Gill,Gill2011} that the UHF energy can capture
part of the static correlation energy and, consequently, the correlation energy $E_{\rm c(U)}$
is more robust than $E_{\rm c(R)}$. In order to better understand the electron correlation,
we will analyze both correlation energy gains
$\Delta E_{\rm c(R)}(Z) = E^{(0)}_{\rm c(R)}(Z) - E^{(+)}_{\rm c(R)}(Z)$
and $\Delta E_{\rm c(U)}(Z) = E^{(0)}_{\rm c(U)}(Z) - E^{(+)}_{\rm c(U)}(Z)$
in the process ``ion + electron → neutral atom”.

Very accurate values of the correlation-energy gains $\Delta E_{\rm c(R)}$ and $\Delta E_{\rm c(U)}$ are
needed in our investigation. The most accurate non-relativistic ground-state energies
$E^{(0)}_{\rm GS}(Z)$ and $E^{(+)}_{\rm GS}(Z)$ (for $Z \le 18$) of the atoms and cations,
respectively, were obtained by Davidson {\it et al.}\cite{Davidson,Davidson93,Davidson96}.
They took an empirical/theoretical
approach and combined experimental ionization energies with corrections to remove
the effect of reduced mass, nuclear shape, and relativistic corrections.
The ground-state energies obtained by Davidson {\it et al.} are widely accepted as the reference
values for atoms and ions with $Z \le 18$. In their work, they also calculated
the RHF energies of the atoms and ions and gave the corresponding correlation energies
$E^{(0)}_{\rm c(R)}(Z)$ and $E^{(+)}_{\rm c(R)}(Z)$. However, there are no such accurate
results available for larger atoms with more than 18 electrons. For example,
the correlation energies calculated by McCarthy and Thakkar\cite{McCarthy11} for atoms
from K to Kr ($Z$=19 to 36) have errors from 22 to 55 m$E_{\rm h}$.
Very recent calculations by Annaberdiyev {\it et al.}\cite{Mitas20} based on the configuration
interaction, coupled cluster, and quantum Monte Carlo methods covering elements up to Kr
have achieved an accuracy of about 1 to 10 m$E_{\rm h}$ for K to Zn atoms ($Z$= 19 to 30).
The difficulty in obtaining accurate correlation energies in many-electron atoms and
molecules is a major obstacle in the study of correlation effects.

\begin{table*}[!htb]
%\centering
%\scalefont{0.9}
\caption{\label{Etotal}
The ground-state energies $E^{(0)}_{\rm GS}(Z)$ and $E^{(+)}_{\rm GS}(Z)$\cite{Davidson96}(for $Z=2$ to 18),
the UHF energies $E^{(0)}_{\rm UHF}(Z)$ and $E^{(+)}_{\rm UHF}(Z)$, and the correlation energies $E^{(0)}_{\rm c(U)}(Z)$
and $E^{(+)}_{\rm c(U)}(Z)$ (in a.u.) for atoms and cations, respectively.
The correlation energy gains $\Delta E_{\rm c(U)}(Z)$ and $\Delta E_{\rm c(R)}(Z)$ are given in unit m$E_{\rm h}$.}
\addtolength{\tabcolsep}{+1pt}
\begin{tabular}{cccccccccccccc}
\hline
\hline
 & & \multicolumn{3}{c}{\text{Atom}} & & & \multicolumn{3}{c}{\text{Cation}} & & \multicolumn{3}{c}{\text{Corr. Energy Gain}}
 (m$E_{\rm h}$) \\
\cline{2-5} \cline{7-10} \cline{12-14}
$Z$&State &$E^{(0)}_{\rm GS}(Z)$ & $E^{(0)}_{\rm UHF}(Z)$ & $E^{(0)}_{\rm c(U)}(Z)$ & & State & $E^{(+)}_{\rm GS}(Z)$
& $E^{(+)}_{\rm UHF}(Z)$ & $E^{(+)}_{\rm c(U)}(Z)$  & & $\Delta E_{\rm c(U)}(Z) $& & $\Delta E_{\rm c(R)}(Z)$  \\
\hline
2  & $^1 S$ & -2.903724& -2.861739& -0.041985 &  &$^2 S$ &  -2.00000&  -2.00000&  0.00000 &	& -42.00 & &-42.04 \\
3  & $^2 S$ & -7.478060& -7.432774& -0.045286 &  &$^1 S$ & -7.279913& -7.236388& -0.043525&	& -1.76	 & &-1.83  \\
4  & $^1 S$ & -14.66739& -14.57302& -0.09437 &   &$^2 S$ & -14.32476& -14.27749& -0.04727&	& -47.10 & &-46.99 \\
5  & $^2 P$ & -24.65390& -24.53319& -0.12071 &   &$^1 S$ & -24.34889& -24.23760& -0.11129&	& -9.42 & &-13.52 \\
6  & $^3 P$ &  -37.8450&  -37.6938& -0.1512 &    &$^2 P$ & -37.43095& -37.29691& -0.1340&	& -17.2	& &-17.63 \\
7  & $^4 S$ &  -54.5893&  -54.4047& -0.1846 &    &$^3 P$ &  -54.0546&  -53.8941& -0.1605&	& -24.1	& &-21.75 \\
8  & $^3 P$ &  -75.0674&  -74.8192& -0.2482 &    &$^4 S$ &  -74.5669&  -74.3773& -0.1896&	& -58.6	& &-63.71 \\
9  & $^2 P$ &  -99.7341&  -99.4166& -0.3175 &    &$^3 P$ &  -99.0930&  -98.8414& -0.2516&	& -65.9	& &-63.50 \\
10 & $^1 S$ & -128.9383& -128.5475& -0.3908 &    &$^2 P$ & -128.1437& -127.8251& -0.3186&	& -72.2	& &-65.36 \\
11 & $^2 S$ & -162.2554& -161.8586& -0.3968 &    &$^1 S$ & -162.0667& -161.6767& -0.3900&   &  -6.8 & &-6.70  \\
12 & $^1 S$ & -200.0540& -199.6150& -0.4390 &    &$^2 S$ & -199.7732& -199.3722& -0.4010&   & -38.0 & &-38.03 \\
13 & $^2 P$ & -242.3470& -241.8809& -0.4661 &    &$^1 S$ & -242.1270& -241.6747& -0.4523&	& -13.8	& &-18.05 \\
14 & $^3 P$ & -289.3600& -288.8590& -0.5010 &    &$^2 P$ & -289.0600& -288.5784& -0.4816&	& -19.4	& &-18.92 \\
15 & $^4 S$ & -341.2600& -340.7194& -0.5406 &    &$^3 P$ & -340.8720& -340.3555& -0.5166&	& -24.0	& &-18.31 \\
16 & $^3 P$ & -398.1110& -397.5136& -0.5974 &    &$^4 S$ & -397.7310& -397.1741& -0.5569&	& -40.5	& &-48.71 \\
17 & $^2 P$ & -460.1500& -459.4902& -0.6599 &    &$^3 P$ & -459.6730& -459.0579& -0.6151&	& -44.8	& &-44.03 \\
18 & $^1 S$ & -527.5440& -526.8177& -0.7263 &    &$^2 P$ & -526.9610& -526.2833& -0.6778&	& -48.5	& &-39.19 \\
\hline\hline
\end{tabular}
\end{table*}

Using the results given by Chakravorty and Davidson\cite{Davidson96} of the non-relativistic ground-state
energies $E_{\rm GS}(Z)$ and the correlation energies $E_{\rm c(R)}(Z)$ of the atoms and cations,
we may obtain the correlation-energy gains for $Z \le 18$. To obtain the correlation
energy $E_{\rm c(U)}(Z)$, we also need the UHF energies of the atoms and cations.
We performed calculations using the cc-pVQZ, cc-pV5Z, and cc-pV6Z basis sets\cite{g16}
to extrapolate to the complete basis set limit\cite{Halkier}. Because the ground-state energies obtained
in Ref.~\onlinecite{Davidson96} for most atoms and ions are with four decimal digits (i.e., they are accurate
to within 0.1 m$E_{\rm h}$), we can determine the correlation energies to an accuracy
in the order of 0.1 m$E_{\rm h}$. The obtained $E^{(0)}_{\rm UHF}(Z)$ and $E^{(+)}_{\rm UHF}(Z)$
are given in Table I together with the ground-state energies $E^{(0)}_{\rm GS}(Z)$ and $E^{(+)}_{\rm GS}(Z)$
from Ref.~\onlinecite{Davidson96}. Their differences are the correlation energies $E^{(0)}_{\rm c(U)}(Z)$
for the atoms and $E^{(+)}_{\rm c(U)}(Z)$ for the ions. The above calculations yield
the correlation-energy gain $\Delta E_{\rm c(U)}(Z) = E^{(0)}_{\rm c(U)}(Z)-E^{(+)}_{\rm c(U)}(Z)$.
The correlation-energy gain $\Delta E_{\rm c(R)}(Z) = E^{(0)}_{\rm c(R)}(Z)-E^{(+)}_{\rm c(R)}(Z)$
can be estimated using the correlation energies $E^{(0)}_{\rm c(R)}(Z)$ and $E^{(+)}_{\rm c(R)}(Z)$
given in Ref.~\onlinecite{Davidson96}. The obtained values of $\Delta E_{\rm c(U)}(Z)$ and $\Delta E_{\rm c(R)}(Z)$
are shown in the last two columns in Table I and they are plotted as a function of $Z$ in Fig. 1.

\begin{figure}[b!]
      {\includegraphics[width=9cm,height=7cm]{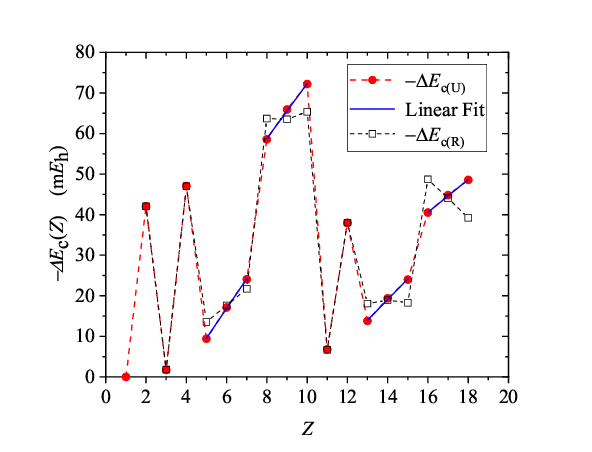}}
       \caption{The correlation-energy gain for $Z\leq 18$.}
       \label{Fig1}
\end{figure}

Fig.~1 shows that both the correlation-energy gains $\Delta E_{\rm c(U)}$ and $\Delta E_{\rm c(R)}$
are large when the added electron occupies an orbital in which there is already another electron
in it. This happens at $Z$ = 2, 4, 8, 9, 10, 12, 16, 17, and 18, and a new electron pair forms
in the same orbital.
Otherwise, the correlation-energy gains are small. The largest difference between the values of
$\Delta E_{\rm c(U)}$ and $\Delta E_{\rm c(R)}$ is not more than 10 m$E_{\rm h}$.
But the most significant difference between them is manifested in their dependence
on the atomic number $Z$. The energy $\Delta E_{\rm c(U)}$ versus $Z$ for $Z$ = 5 to 7, as well
as for $Z$ = 8 to 10 (which are associated with the 2$p$ valence electrons) shows very good linear relations.
The same is happening for the correlation-energy gain $\Delta E_{\rm c(U)}$ associated with
the 3$p$ valence electrons for $Z$ = 13 to 18. But $\Delta E_{\rm c(R)}(Z)$ does not show
such a linear behavior for $p$ electrons. The piecewise linear dependence of the energy $\Delta E_{\rm c(U)}(Z)$
on $Z$ reveals that the inter-orbital and intra-orbital correlation energies
associated with the valence electrons are not interdependent.
But it occurs only for the correlation energy
$E_{\rm c(U)}$ beyond the UHF energy defined by Pople and Binkley\cite{Pople} due mostly to
the dynamical correlation. Therefore, we will focus on the correlation-energy gains
$\Delta E_{\rm c(U)}$ for the valence electrons to show the regularities of
the electron correlation energy associated with the valence electrons in atoms.

When a helium cation binds one more electron becoming a neutral atom,
the correlation-energy gain $\Delta E_c(2)$ is determined by the intra-orbital PCE
between two electrons occupying the same 1$s$ orbital ($\varepsilon_{1s^2}$) of the He atom. From Table I
we obtain $\Delta E_{\rm c(U)}(2) = -42.00$ m$E_{\rm h} = \varepsilon_{1s^2}$.
For a lithium cation, there are two electrons in the core 1$s$ orbital with correlation energy
$E^{(+)}_{\rm c}(3)= \varepsilon_{{\rm core}(1s^2)}$. When it catches one more electron becoming a neutral Li atom,
the third electron occupies the 2$s$ orbital and the total correlation energy of the Li atom is given by
$E^{(0)}_{\rm c}(3)= \varepsilon_{{\rm core}(1s^2)} + \varepsilon_{2s\mbox{-}1s^2}$ with the correlation-energy gain
$\Delta E_{\rm c}(3)= \varepsilon_{2s\mbox{-}1s^2}$.
In this case, the influence of the third electron on the core-electron correlation in the 1$s$ orbital is very small
and can be ignored. The correlation-energy gain $\Delta E_{\rm c}(3)$ is due to the inter-shell correlation energy
$\varepsilon_{2s\mbox{-}1s^2} \equiv \varepsilon_{2s}^{\rm I}$ between one valence electron in the 2$s$ orbital
and two core-electrons in the 1$s$ orbital in the K shell. Using the value of $\Delta E_{\rm c(U)}(3)$ we obtain
$\varepsilon_{2s}^{\rm I} = \Delta E_{\rm c(U)}(3) = -1.76$ m$E_{\rm h}$.

Similarly, adding one electron to a beryllium cation the correlation-energy gain from the 3-electron cation
to 4-electron neutral atom is given by $\Delta E_{\rm c}(4)
=\varepsilon_{2s^2\mbox{-}1s^2}-\varepsilon_{2s}^{\rm I} +\varepsilon_{2s^2}$, where $\varepsilon_{2s^2\mbox{-}1s^2}$
is the inter-shell correlation energy between the two 2$s$ valence electrons and two 1$s$ core electrons
and $\varepsilon_{2s^2}$ is the intra-orbital PCE between the two electrons in the same 2$s$ orbital.
The energy $\varepsilon_{2s^2\mbox{-}1s^2}$ should be different from 2$\varepsilon_{2s}^{\rm I}$
because the one electron density distribution in a singly occupied 2$s$ orbital is different
from that in the doubly occupied. We may introduce a constant $\eta_{2s}$ and denote
$\varepsilon_{2s^2\mbox{-}1s^2}= 2\eta_{2s} \varepsilon_{2s}^{\rm I}$.
Then the correlation-energy gain is given by
$\Delta E_{\rm c}(4) = \varepsilon_{2s^\prime}^{\rm I}+\varepsilon_{2s^2}$ with
$\varepsilon_{2s^\prime}^{\rm I}= (2\eta_{2s}-1)\varepsilon_{2s}^{\rm I}$.

For $Z=2$ to 7, we can obtain the following expressions for the correlation-energy gain,
\begin{eqnarray}
&& \Delta E_c(2)= \varepsilon_{1s^2}, \label{DEc2} \\
&& \Delta E_c(3)= \varepsilon_{2s\mbox{-}1s^2} \equiv \varepsilon_{2s}^{\rm I}, \label{DEc3} \\
&& \Delta E_c(4)=  \varepsilon_{2s^\prime}^{\rm I} + \varepsilon_{2s^2}, \label{DEc4} \\
&& \Delta E_c(5)=  \varepsilon_{2p}^{\rm II},\label{DEc5} \\
&& \Delta E_c(6)= \varepsilon_{2p}^{\rm II} + \varepsilon_{2p\mbox{-}2p}, \label{DEc6} \\
&& \Delta E_c(7)= \varepsilon_{2p}^{\rm II} + 2\varepsilon_{2p\mbox{-}2p}. \label{DEc7}
\end{eqnarray}
Eq.~(\ref{DEc5}) shows that $\Delta E_c(5)$ for boron atom is due to the (inter-shell-inter-subshell) correlation energy
$\varepsilon_{2p}^{\rm II} \equiv \varepsilon_{2p\mbox{-}(2s^2 1s^2)}$ between one 2$p$ valence electron and the rest four
in the 2$s$ subshell and 1$s$ core. In carbon atom ($Z$=6), two 2$p$ electrons occupy different orbitals by Hund's rule,
being with the inter-orbital PCE $\varepsilon_{2p\mbox{-}2p}$ between them.
Thus $\Delta E_c(6)$ is given by $\varepsilon_{2p}^{\rm II}$ for the added electron and the inter-orbital
PCE $\varepsilon_{2p\mbox{-}2p}$.
Eqs.~(\ref{DEc5}-\ref{DEc7}) shows that $\Delta E_c(Z)$ has a linear dependence on $Z$ for
$Z=$5 to 7 with the slope $\varepsilon^{(2p)}_{\rm slope}=\varepsilon_{2p\mbox{-}2p}$.
It means that the inter-orbital correlation energies $\varepsilon_{2p\mbox{-}2p}$ and
$\varepsilon_{2p}^{\rm II}$ for 2$p$ valence electrons in B, C, and N atoms should have
the same values. The best linear fitting to the values of $\Delta E_c(Z)$ versus $Z$ for $Z=$5 to 7
in Fig. 1 yields the slope $\varepsilon^{(2p)}_{\rm slope}=\varepsilon_{2p\mbox{-}2p}=-7.3\pm 0.2$
m$E_{\rm h}$ and the intercept $\varepsilon_{2p}^{\rm II}=-9.6\pm 0.3$ m$E_{\rm h}$,
with the standard errors of a few tenths millihartree. Therefore, our analysis
is confirmed by the best numerical results of the correlation energies in atoms and,
within the accuracy of a few tenths m$E_{\rm h}$, the inter-orbital correlation energies
$\varepsilon_{2p\mbox{-}2p}$ and $\varepsilon_{2p}^{\rm II}$ are not
dependent on each other.

In the ground state of oxygen atom ($Z$=8), there is a doubly occupied 2$p$ orbital where
two electrons of opposite spins occupying the same orbital. The corresponding correlation-energy gain is given by
\[
\Delta E_c(8)= \varepsilon_{2p^2\mbox{-}(2s^2 1s^2)} -\varepsilon_{2p}^{\rm II} + 2 \varepsilon_{2p^2\mbox{-}2p}
- 2 \varepsilon_{2p\mbox{-}2p} + \varepsilon_{2p^2}.
\]
Because the charge distribution for each electron in doubly occupied 2$p$ orbital is different from
that of the singly occupied one, in principle it leads to $\varepsilon_{2p^2\mbox{-}2p} \ne 2 \varepsilon_{2p\mbox{-}2p}$.
Thus, we introduce a dimensionless parameter $\eta_{2p}$ such that
$\varepsilon_{2p^2\mbox{-}2p}= 2 \varepsilon^\prime_{2p\mbox{-}2p}$ with
$\varepsilon^\prime_{2p\mbox{-}2p} = \eta_{2p} \varepsilon_{2p\mbox{-}2p}$ and also
$\varepsilon_{2p^2\mbox{-}(2s^2 1s^2)}=2\varepsilon^\prime_{2p \mbox{-}(2s^2 1s^2)}$ with
$\varepsilon^\prime_{2p \mbox{-}(2s^2 1s^2)} =  \eta_{2p}\varepsilon_{2p}^{\rm II}$
to quantify this alteration with $\eta_{2p}$ a constant of value close to but different from 1.
Then, the expression for $\Delta E_c(8)$ becomes $\Delta E_c(8)= (2\eta_{2p}-1)(\varepsilon_{2p}^{\rm II}
+2\varepsilon_{2p\mbox{-}2p}) + \varepsilon_{2p^2} \equiv \varepsilon^{\rm III}_{2p}$.

For fluorine ($Z$=9) and neon ($Z$=10) atoms with two and three doubly occupied 2$p$ orbitals,
respectively, the inter-orbital correlation energy $\varepsilon_{2p^2\mbox{-}2p^2}$ between
two 2$p$ electrons in one orbital and the other two in another 2$p$ orbital appears.
Using the parameter $\eta_{2p}$, we have $\varepsilon_{2p^2\mbox{-}2p^2}
= 2 \eta_{2p}\varepsilon_{2p^2\mbox{-}2p} = 4 \eta^2_{2p}\varepsilon_{2p\mbox{-}2p}$.
The expressions for $\Delta E_c(Z)$ can be written as, for $Z=8$, 9 and 10,
\begin{eqnarray}
&& \Delta E_c(8)= \varepsilon^{\rm III}_{2p}, \label{DEc8} \\
&& \Delta E_c(9)= \varepsilon^{\rm III}_{2p}+ (2\eta_{2p}-1)^2 \varepsilon_{2p\mbox{-}2p},\label{DEc9}\\
&& \Delta E_c(10)= \varepsilon^{\rm III}_{2p}+ 2 (2\eta_{2p}-1)^2\varepsilon_{2p\mbox{-}2p}. \label{DEc10}
\end{eqnarray}
We see that $\Delta E_c(Z)$ versus $Z$ for $Z=$8 to 10 also has a linear relation but with
the slope $\varepsilon^{(2p)}_{\rm slope^\prime}= (2\eta_{2p}-1)^2 \varepsilon_{2p\mbox{-}2p}$.
Therefore, the valence electrons in doubly occupied 2$p$ orbitals in O, F and Ne atoms also share
the same values of the intra- and inter-orbital pair-correlation energies.
The linear regression of $\Delta E_{\rm c(U)}(Z)$ for $Z=$8 to 10 in Fig.~1 yields the slope
$\varepsilon^{(2p)}_{\rm slope^\prime}=(2\eta_{2p}-1)^2 \varepsilon_{2p\mbox{-}2p}=-6.8\pm 0.3$
m$E_{\rm h}$ and the intercept $\varepsilon^{\rm III}_{2p}=-58.8\pm 0.4$ m$E_{\rm h}$.
Again, we observe that the standard errors for both the slope and intercept obtained from
the fitting are a few tenths millihartree. From the above results, we obtain $\eta_{2p}=0.982\pm 0.013$ and
$\varepsilon_{2p^2}=-35.5\pm 0.8$  m$E_{\rm h}$. Thus, the inter-orbital PCE $\varepsilon^\prime_{2p\mbox{-}2p}$
is given by $\varepsilon^\prime_{2p\mbox{-}2p}=\eta_{2p} \varepsilon_{2p\mbox{-}2p} = -7.2\pm 0.2$ m$E_{\rm h}$
which is only about 2\% smaller than $\varepsilon_{2p\mbox{-}2p}$ due to double occupation of a 2$p$ orbital.
It demonstrates that the strong direct Coulomb repulsion between two electrons in the same orbital
affects very slightly the related inter-orbital PCEs.

It is known that there is orbital relaxation when we take one electron out from the atom or
add one electron into the cation. As a matter of fact, the most significant orbital
relation occurs when we remove one electron from a doubly occupied orbital or
when we add one electron into an orbital where there is already one electron in it.
It is generally believed that strong Coulomb repulsion between two electrons occupying the same orbital
alters the electron density distribution and induces strong orbital relaxation effects on
the correlation energy. Here, we show the difference
$\varepsilon_{2p\mbox{-}2p}-\varepsilon^\prime_{2p\mbox{-}2p}=(1- \eta_{2p}) \varepsilon_{2p\mbox{-}2p}
\simeq 0.1\pm 0.2$ m$E_{\rm h}$ is quantitatively responsible for the change of inter-orbital PCE result from
orbital relaxation. This is the most important ingredient for the correlation energy change
due to orbital relaxation. We find that the effects of the orbital relaxation on the correlation
energy are surprisingly small, contributing less that 2\%  change of the correlation energy in this case.

The above results indicate clearly that the inter-orbital
PCEs $\varepsilon_{2p\mbox{-}2p}$ and $\varepsilon^\prime_{2p\mbox{-}2p}$ do not intertwine
with the intra-orbital PCE $\varepsilon_{2p^2}$. The inter-orbital PCEs are also independent of the spins
of the electrons because two electrons with the same or opposite spins have the same value of inter-orbital PCE.
Notice that the intra-orbital PCE in atoms is always for two electrons of opposite spins.
%{\it It also shows that the correlation energies
%in atoms are basically determined by two-body interactions and the multibody correlation
%effects consisting of three- and four-body terms are very small.}

We now come back to exam the correlation energy gain $\Delta E_{\rm c(U)}(4)$ given by Eq.~(\ref{DEc4})
and its components $\varepsilon_{2s^\prime}^{\rm I}= (2\eta_{2s}-1)\varepsilon_{2s}^{\rm I}$ and $\varepsilon_{2s^2}$.
Because the present analysis cannot provide exact value for $\eta_{2s}$, considering the result obtained
above for $\eta_{2p}$ we may assume that the double occupation of the 2$s$ orbital affects very little
the related inter-orbital correlation energy and approximate $\eta_{2s}\simeq 1$, i.e.,
$\varepsilon_{2s^\prime}^{\rm I} \simeq \varepsilon_{2s}^{\rm I}= -1.76$ m$E_{\rm h}$.
Using the value $\Delta E_{\rm c(U)}(4) = -47.10$ m$E_{\rm h}$ we obtain
$\varepsilon_{2s^2} \simeq -45.3$ m$E_{\rm h}$.

For the atoms in the third row of the periodic table, following the similar discussions and notations
used above for the atoms in the second row, the correlation-energy gains $\Delta E_c(Z)$ for $Z=11$ to 18 can be written as,
\begin{eqnarray}
&& \Delta E_c(11)= \varepsilon_{3s\mbox{-}{\rm (core)}} \equiv \varepsilon_{3s}^{\rm I}, \label{DEc11} \\
&& \Delta E_c(12)= \varepsilon_{3s^\prime}^{\rm I} + \varepsilon_{3s^2}, \label{DEc12} \\
&& \Delta E_c(13)= \varepsilon_{3p\mbox{-}(3s^2{\rm core})} \equiv \varepsilon_{3p}^{\rm II},\label{DEc13} \\
&& \Delta E_c(14)= \varepsilon_{3p}^{\rm II} + \varepsilon_{3p\mbox{-}3p}, \label{DEc14} \\
&& \Delta E_c(15)= \varepsilon_{3p}^{\rm II} + 2 \varepsilon_{3p\mbox{-}3p}, \label{DEc15} \\
&& \Delta E_c(16)=  \varepsilon^{\rm III}_{3p}, \label{DEc16} \\
&& \Delta E_c(17)=  \varepsilon^{\rm III}_{3p}+ (2\eta_{3p}-1)^2 \varepsilon_{3p\mbox{-}3p} , \label{DEc17}\\
&& \Delta E_c(18)= \varepsilon^{\rm III}_{3p}+ 2 (2\eta_{3p}-1)^2\varepsilon_{3p\mbox{-}3p}. \label{DEc18}
\end{eqnarray}
In Eq.~(\ref{DEc11}), $\varepsilon_{3s}^{\rm I}$ is the inter-shell correlation energy between a 3$s$
valence electron and all the core electrons in the K and L shells.
From Table I we obtain $\varepsilon_{3s}^{\rm I} =\Delta E_{\rm c(U)}(11)= -6.8$ m$E_{\rm h}$. Being similar to
Eq.~(\ref{DEc4}), $\Delta E_c(12)$ in Eq.~(\ref{DEc12}) is a sum of $\varepsilon_{3s^\prime}^{\rm I}$
and the intra-orbital PCE $\varepsilon_{3s^2}$.
Assuming $\varepsilon_{3s^\prime}^{\rm I} \simeq \varepsilon_{3s}^{\rm I}$
and using the value $\Delta E_{\rm c(U)}(12)=-38.0$ m$E_{\rm h}$, we obtain $\varepsilon_{3s^2} \simeq 31.2$ m$E_{\rm h}$.
Based on the relation indicated by Eqs.~(\ref{DEc13}-\ref{DEc15}), a linear regression to the values of $\Delta E_{\rm c(U)}(Z)$
vs. $Z$ (for $Z=13$ to 15) in Fig.~1 yields the slope $\varepsilon^{(\rm 3p)}_{\rm slope}
=\varepsilon_{3p\mbox{-}3p}=-5.1\pm 0.2$ m$E_{\rm h}$
and the intercept $\varepsilon_{3p}^{\rm II}=-14.0\pm 0.3$ m$E_{\rm h}$, where $\varepsilon_{3p}^{\rm II}$ is
the correlation energy between a 3$p$ valence electron and the rest in the 3$s$ subshell and in the core (K and L shells).
To characterize the difference in the inter-orbital PCE for singly and doubly occupied 3$p$ orbitals,
we follow the technique used for 2$p$ electrons and introduce the parameter $\eta_{3p}$ such that
$\varepsilon_{3p^2\mbox{-}3p}= 2 \varepsilon^{\prime}_{3p\mbox{-}3p}$ with
$\varepsilon^{\prime}_{3p\mbox{-}3p} = \eta_{3p} \varepsilon_{3p\mbox{-}3p}$ and
$\varepsilon_{3p^2\mbox{-}(3s^2{\rm core})}=2 \eta_{3p}\varepsilon_{3p}^{\rm II}$.
In this way, the quantity $\varepsilon^{\rm III}_{3p}$ appeared in Eqs.~(\ref{DEc16}-\ref{DEc18}) is given by
$\varepsilon^{\rm III}_{3p} \equiv (2\eta_{3p}-1)(\varepsilon_{3p}^{\rm II} +2\varepsilon_{3p\mbox{-}3p}) + \varepsilon_{3p^2}$.
The best linear fitting to the values of $\Delta E_{\rm c(U)}(Z)$ vs. $Z$ (for $Z=$16 to 18) in Fig.~1 yields the slope
$\varepsilon^{(3p)}_{\rm slope^\prime} =(2\eta_{3p}-1)^2 \varepsilon_{3p\mbox{-}3p} =-4.0\pm 0.1$ m$E_{\rm h}$
and the intercept $\varepsilon^{\rm III}_{3p}=-40.6\pm 0.2$ m$E_{\rm h}$.
Using the above results from the linear fitting, we obtain $\eta_{3p}=0.943\pm 0.010$,
$\varepsilon^{\prime}_{3p\mbox{-}3p}=4.8\pm 0.2$ m$E_{\rm h}$, and $\varepsilon_{3p^2}=-19.2\pm 0.7$ m$E_{\rm h}$.
We see that the difference between $\varepsilon_{3p\mbox{-}3p}$ and $\varepsilon^{\prime}_{3p\mbox{-}3p}$
due to orbital relation
is larger than that between $\varepsilon_{2p\mbox{-}2p}$ and $\varepsilon^\prime_{2p\mbox{-}2p}$, but it
is still quite small being close to the standard error from the fitting. The above results show that for
the valence electrons in the third period, the inter-orbital and intra-orbital correlation energies
exhibit similar behaviors to those of the second period.

Notice that $\varepsilon_{2s}^{\rm I}$ ($\varepsilon_{3s}^{\rm I}$) is the correlation energy between 1 valence electron
in $2s$ ($3s$) orbital and 2 (10) core electrons. And $\varepsilon_{2p}^{\rm II}$ ($\varepsilon_{3p}^{\rm II}$) is
that between 1 valence electron in $2p$ ($3p$) orbital and the rest ones in the $2s$ ($3s$) subshell and the ionic core.
It is reasonable to assume that the 2$s$ and 2$p$ (3$s$ and 3$p$) valence electrons have the same inter-shell
correlation energy with the core electrons, i.e.,
$\varepsilon_{2p\mbox{-}1s^2} =\varepsilon_{2s\mbox{-}1s^2}\equiv \varepsilon_{2s}^{\rm I} $
($\varepsilon_{3p\mbox{-}{\rm core}} =\varepsilon_{3s\mbox{-}{\rm core}}\equiv \varepsilon_{3s}^{\rm I}$ ).
Within such approximations, we obtain the inter-subshell PCEs
$\varepsilon_{2p\mbox{-}2s} = -3.9\pm 0.2$ m$E_{\rm h}$ and $\varepsilon_{3p\mbox{-}3s} \simeq -3.6\pm 0.2 $ m$E_{\rm h}$.
Consequently, the average inter-shell pair-correlation energy between a 2$s$ or a 2$p$ valence electron
and a core electron is given by $\bar{\varepsilon}_{2}^{\rm I}= \varepsilon_{2s}^{\rm I}/2 = -0.88$ m$E_{\rm h}$.
And that between a 3$s$ or a 3$p$ valence electron and a core electron is given by
$\bar{\varepsilon}_{3}^{\rm I}= \varepsilon_{3s}^{\rm I}/10 = -0.68$ m$E_{\rm h}$.

In summary, we have investigated and compared the different correlation energies
$E_{\rm c(R)}$ and $E_{\rm c(U)}$ of atoms based on the most accurate ground-state energies
for $Z \le 18$. We show that the correlation-energy gain $\Delta E_{\rm c(U)}(Z)$ associated
with the valence electrons in atoms is given by a linear combination of the intra- and inter-orbital PCEs.
We reveal that within the chemical accuracy, the PECs associated with the valence
electrons of the atoms in the same row of the periodic table have the same values.
These PCEs are not entangled and their values depend only on the electron orbitals.
For two specific orbitals, the inter-orbital correlation energy is
the same for two electrons of parallel spins or anti-parallel spins.
Furthermore, We find that the effects of orbital relaxation on the correlation energy are very small.

\begin{table}[t!]
\centering
\caption{The intra-orbital, inter-orbital, and the average inter-shell PCEs (in unit m$E_{\rm h}$)
associated with the valence electrons in atoms in the first three rows of the periodic table.}
\begin{tabular}{cccccccc}
\hline
\hline
 \multicolumn{2}{c}{Intra-orbital PCE}& &\multicolumn{2}{c}{Inter-orbital PCE}& &\multicolumn{2}{c}{Inter-shell PCE} \\
\cline{1-2} \cline{4-5} \cline{7-8}
$\varepsilon_{1s^2}$ & -42.0        \\
$\varepsilon_{2s^2}$ & -45.3 & &  $\varepsilon_{2p\mbox{-}2s}$ & \;\;  -3.9  & \;\;  &   &   \\
$\varepsilon_{2p^2}$ & -35.5 & \;\;\;  & $\varepsilon_{2p\mbox{-}2p}$ & \;\; -7.3 & \;\;\;
& $\bar{\varepsilon}_{2}^{\rm I}$&\;\;\; -0.88 \\
                     &       & \;\;\;  & $\varepsilon^{\prime}_{2p\mbox{-}2p}$ & \;\; -7.2 & \;\;\;
& &\;\;\; \\
$\varepsilon_{3s^2}$ & -31.2   & \;\;\;  & $\varepsilon_{3p\mbox{-}3s}$ & \;\; -3.6 &  \;\; & \\
$\varepsilon_{3p^2}$ & -19.2 &  & $\varepsilon_{3p\mbox{-}3p}$ & \;\; -5.1 & & $\bar{\varepsilon}_{3}^{\rm I}$& \;\;\; -0.68 \\
                     &       & \;\;\;  & $\varepsilon^{\prime}_{3p\mbox{-}3p}$ & \;\; -4.8 & \;\;\;
& &\;\;\;
\\[1ex]
\hline\hline
\end{tabular}
\end{table}

The obtained values of the intra-orbital, inter-orbital, and average inter-shell PCEs for
the atoms in the first three rows of the periodic table are given in Table II.
Using the obtained PCEs in this table, the correlation energy gains $\Delta E_{\rm c(U)}(Z)$
for the valence electrons shown in Table I can be recovered according to Eqs. (1-17).
The PCEs associated with the valence electrons in 2$p$ orbitals in B, C, N, O, F and Ne atoms
in the second row have the same value $\varepsilon_{2p^2}$, $\varepsilon_{2p\mbox{-}2p}$,
$\varepsilon^{\prime}_{2p\mbox{-}2p}$ and $\varepsilon_{2p\mbox{-}2s}$ given in Table II,
and those associated with the valence electrons in 3$p$ orbitals in Al, Si, P, S, Cl and Ar atoms
in the third row have the same PCEs $\varepsilon_{3p^2}$, $\varepsilon_{3p\mbox{-}3p}$,
$\varepsilon^{\prime}_{3p\mbox{-}3p}$ and $\varepsilon_{3p\mbox{-}3s}$, etc.
Quantitatively, the intra-orbital PCEs are much larger than the inter-orbital
and inter-shell ones. The inter-orbital PCE for electrons in the same shell is about one order
of magnitude smaller than the intra-orbital one. The inter-shell PCE between a valence
and a core electron is even smaller. The distinction between the intra-orbital and inter-orbital
correlation may have important implications because they are not entangled and
the intra-orbital PCE is much larger. Two electrons occupying the same orbital
may lose their single-particle identity within the correlation effects acting as a new
entity of a bound electron pair, which should manifest itself
in the properties of some molecules and materials.\cite{Hai18,Pina,Hai24}

\section{Acknowledgments}
This research was supported by FAPESP (S{\~a}o Paulo Research Foundation, under the grant No. 2024/00484-2),
FAPEMIG, FAPMG, and CNPq (Brazil). The Authors appreciate helpful
discussions with Tomaz Catunda.

\bibliography{Corr_Atoms_Hai.tex}% Produces the bibliography via BibTeX.

\end{document}